\let\ifarxiv=\iftrue     
\pdfoutput=1
{\def\usepackage{ws-procs9x6}}

\ifarxiv

\documentclass[12pt,a4paper]{article}
\usepackage[a4paper,text={450pt,650pt},centering]{geometry}

\fi

\ifarxiv\else

\documentclass[11pt,a4paper]{article}
\usepackage{mathptmx}
\usepackage[a4paper,text={130mm,198mm}]{geometry}

\fi

\ifnum\pdfoutput=1\else
\PassOptionsToPackage{hypertex}{hyperref}
\PassOptionsToPackage{draft}{graphicx}
\usepackage{showkeys}
\fi

\usepackage{amsmath,amssymb}
\usepackage[bookmarks=true,hyperfigures=true]{hyperref}
\usepackage{graphicx}
\usepackage[nosort]{cite}
\ifarxiv\usepackage[bulletsep]{collref}\fi
\usepackage{wrapfig}

\let\oldbfseries=\bfseries
\let\oldmdseries=\mdseries
\let\oldnormalfont=\normalfont
\renewcommand{\bfseries}{\oldbfseries\boldmath}
\renewcommand{\mdseries}{\oldmdseries\unboldmath}
\renewcommand{\normalfont}{\oldnormalfont\unboldmath}

\allowdisplaybreaks[3]

\numberwithin{equation}{section}

\usepackage[font=small,labelfont=bf,width=0.85\textwidth]{caption}

\providecommand{\hypersetup}[1]{}
\providecommand{\texorpdfstring}[2]{#1}

\hypersetup{plainpages=false}
\hypersetup{pdfpagemode=UseNone}
\hypersetup{bookmarksnumbered=true}
\hypersetup{pdfstartview=FitH}
\hypersetup{colorlinks=false}
\hypersetup{citebordercolor={.5 1 .5}}
\hypersetup{urlbordercolor={.5 1 1}}
\hypersetup{linkbordercolor={1 .7 .7}}

\DeclareMathSymbol{\Gamma}{\mathalpha}{letters}{"00}
\DeclareMathSymbol{\Delta}{\mathalpha}{letters}{"01}
\DeclareMathSymbol{\Theta}{\mathalpha}{letters}{"02}
\DeclareMathSymbol{\Lambda}{\mathalpha}{letters}{"03}
\DeclareMathSymbol{\Xi}{\mathalpha}{letters}{"04}
\DeclareMathSymbol{\Pi}{\mathalpha}{letters}{"05}
\DeclareMathSymbol{\Sigma}{\mathalpha}{letters}{"06}
\DeclareMathSymbol{\Upsilon}{\mathalpha}{letters}{"07}
\DeclareMathSymbol{\Phi}{\mathalpha}{letters}{"08}
\DeclareMathSymbol{\Psi}{\mathalpha}{letters}{"09}
\DeclareMathSymbol{\Omega}{\mathalpha}{letters}{"0A}

\newcommand{\superN}{\mathcal{N}}

\newcommand{\Tr}{\mathop{\mathrm{Tr}}}
\newcommand{\STr}{\mathop{\mathrm{STr}}}

\newcommand{\Complex}{\mathbb{C}}

\ifx\genfrac\sdflkaj\else\fi
\newcommand{\sfrac}[2]{{\textstyle\frac{#1}{#2}}}
\newcommand{\half}{\sfrac{1}{2}}
\newcommand{\ihalf}{\sfrac{i}{2}}
\newcommand{\quarter}{\sfrac{1}{4}}

\newcommand{\indup}[1]{_{\mathrm{#1}}}

\newcommand{\supup}[1]{^{\mathrm{#1}}}

\newcommand{\matr}[2]{\left(\begin{array}{#1}#2\end{array}\right)}

\newcommand{\lrbrk}[1]{\left(#1\right)}
\newcommand{\bigbrk}[1]{\bigl(#1\bigr)}

\newcommand{\alg}[1]{\mathrm{#1}}
\newcommand{\grp}[1]{\mathrm{#1}}

\makeatletter
\def\mr@ignsp#1 {\ifx\:#1\@empty\else #1\expandafter\mr@ignsp\fi}%
\newcommand{\multiref}[1]{\begingroup
\xdef\mr@no@sparg{\expandafter\mr@ignsp#1 \: }%
\def\mr@comma{}%
\@for\mr@refs:=\mr@no@sparg\do{\mr@comma\def\mr@comma{,}\ref{\mr@refs}}%
\endgroup}
\makeatother

\renewcommand{\eqref}[1]{(\multiref{#1})}

\newcommand{\hypref}[2]{\ifx\hyperref\asklfhas #2\else\hyperref[#1]{#2}\fi}
\newcommand{\namedref}[2]{\hypref{#2}{#1~\ref*{#2}}}

\newcommand{\figref}[1]{\namedref{Fig.}{#1}}

\ifarxiv
\newcommand{\chapref}[1]{\namedref{Chapter}{chap#1}}
\else
\newcommand{\chapref}[1]{Chapter~\ref*{chap:#1} \cite{chap#1}}
\fi

\providecommand{\arxivref}[2]{\href{http://arxiv.org/abs/#1}{#2}}
\providecommand{\doiref}[2]{\href{http://dx.doi.org/#1}{#2}}

\providecommand{\href}[2]{#2}
\providecommand{\arxivlink}[1]{\href{http://arxiv.org/abs/#1}{arxiv:#1}}

\makeatletter
\newlength{\apb@width}
\newcommand{\autoparbox}[2][c]{\settowidth{\apb@width}{#2}\parbox[#1]{\apb@width}{#2}}
\newcommand{\includegraphicsbox}[2][]{\autoparbox{\includegraphics[#1]{#2}}}
\makeatother

\def\figinsertoneline{\ifarxiv15pt\else15pt\fi}
\def\figinserttwoline{\ifarxiv34pt\else30pt\fi}
\newcommand{\figinsert}[2][\figinsertoneline]{%
\columnsep0.5em%
\begin{wrapfigure}{r}{3cm}\raggedleft\vspace{-\intextsep}\vspace{-1.5ex}\vspace{-#1}%
\includegraphics[width=3cm]{#2}\vspace{-0ex}\vspace{-\intextsep}%
\end{wrapfigure}%
\ignorespaces}

\newcommand{\figinsertp}[2][\ifarxiv25pt\else20pt\fi]{%
\columnsep0.5em%
\begin{wrapfigure}{r}{\ifarxiv3.2cm\else3.5cm\fi}\raggedleft\vspace{-\intextsep}\vspace{-2.3ex}\vspace{0.5ex}\vspace{-#1}%
\includegraphics[width=\ifarxiv3.2cm\else3.5cm\fi]{#2}\vspace{-0ex}\vspace{-\intextsep}%
\end{wrapfigure}%
\ignorespaces}

\newdimen\inslwid
\newcommand{\nextparspace}[1][3.0cm]{%
\setlength{\inslwid}{\textwidth}%
\addtolength{\inslwid}{-#1}%
\addtolength{\inslwid}{-0.5cm}%
\everypar{\parshape 1 0pt \inslwid\everypar{}}%
\ignorespaces}

\begin{document}


\thispagestyle{empty}
\phantomsection
\addcontentsline{toc}{section}{Title}

\begin{flushright}\footnotesize%
\texttt{\arxivlink{1012.3982}},
\texttt{AEI-2010-175}\\
\texttt{CERN-PH-TH/2010-306},
\texttt{HU-EP-10/87}\\
\texttt{HU-MATH-2010-22}, 
\texttt{kcl-mth-10-10}\\
\texttt{UMTG-270},
\texttt{UUITP-41/10}%
\vspace{1em}%
\end{flushright}

\begingroup\parindent0pt
\begingroup\bfseries\ifarxiv\Large\else\LARGE\fi
\hypersetup{pdftitle={Review of AdS/CFT Integrability: An Overview}}%
Review of AdS/CFT Integrability:\\
An Overview
\par\endgroup
\vspace{1.5em}
\begingroup\raggedright\ifarxiv\scshape\else\large\fi%
\hypersetup{pdfauthor={Niklas Beisert et al.}}%
Niklas~Beisert$^{\dagger,\ref{InstAEI}}$, 
Changrim~Ahn$^{\ref{InstEwha}}$,
Luis~F.~Alday$^{\ref{InstOxford},\ref{InstIAS}}$,
Zolt\'an~Bajnok$^{\ref{InstBudapest}}$,
James~M.~Drummond$^{\ref{InstCERN},\ref{InstLAPTH}}$,
Lisa~Freyhult$^{\ref{InstUppsala}}$,
Nikolay~Gromov$^{\ref{InstKings},\ref{InstPNPI}}$,
Romuald~A.~Janik$^{\ref{InstKrakow}}$,
Vladimir~Kazakov$^{\ref{InstENSP},\ref{InstParis6}}$,
Thomas~Klose$^{\ref{InstUppsala},\ref{InstPCTP}}$,
Gregory~P.~Korchemsky$^{\ref{InstSaclay}}$,
Charlotte~Kristjansen$^{\ref{InstNBI}}$,
Marc~Magro$^{\ref{InstENSL},\ref{InstAEI}}$,
Tristan~McLoughlin$^{\ref{InstAEI}}$,
Joseph~A.~Minahan$^{\ref{InstUppsala}}$,
Rafael~I.~Nepomechie$^{\ref{InstFlorida}}$,
Adam~Rej$^{\ref{InstIC}}$,
Radu~Roiban$^{\ref{InstPSU}}$,
Sakura~Sch{\"a}fer-Nameki$^{\ref{InstKings},\ref{InstKITP}}$,
Christoph~Sieg$^{\ref{InstHU},\ref{InstNBIA}}$,
Matthias~Staudacher$^{\ref{InstHU},\ref{InstAEI}}$,
Alessandro~Torrielli$^{\ref{InstYork},\ref{InstUtrecht}}$,
Arkady~A.~Tseytlin$^{\ref{InstIC}}$,
Pedro~Vieira$^{\ref{InstPITP}}$,
Dmytro~Volin$^{\ref{InstPSU}}$ 
  and
Konstantinos~Zoubos$^{\ref{InstNBI}}$
\par\endgroup
\vspace{1em}
\begingroup
$^\dagger$%
corresponding author, e-mail address:
\ttfamily
nbeisert@aei.mpg.de
\par\endgroup
\vspace{1.0em}
\endgroup

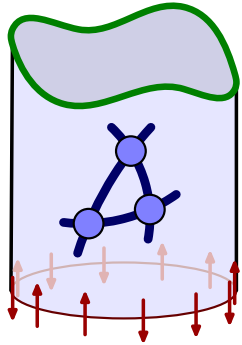
\begin{center}
\includegraphics[width=5cm]{FigTitle.mps}
\vspace{1.0em}
\end{center}

\paragraph{Abstract:}
This is the introductory chapter of a review collection on
integrability in the context of the AdS/CFT correspondence. 
In the collection we present an overview of the achievements 
and the status of this subject as of the year 2010.

\ifarxiv\else
\fi
\hypersetup{pdfsubject={MSC (2010): 37K15, 81T13, 81T30, 81U15}}%

\ifarxiv\else
\fi
\hypersetup{pdfkeywords={gauge theory, string theory, duality, integrability, Bethe ansatz}}%

\newpage

\paragraph{Addresses:}

\begingroup\itshape
\newcounter{instno}
\begin{list}{$^{\arabic{instno}}$}{
  \settowidth{\labelwidth}{$^{10}$}
  \itemsep0pt
  \parsep0.2ex
  \leftmargin\labelwidth
  \addtolength{\leftmargin}{\labelsep}
  \usecounter{instno}
  \small
}

\item\label{InstAEI}
Max-Planck-Institut f\"ur Gravitationsphysik,
Albert-Einstein-Institut\\
Am M\"uhlenberg 1,
14476 Potsdam, 
Germany

\item\label{InstEwha}
Department of Physics and Institute for the Early Universe\\
Ewha Womans University, 
Seoul 120-750, 
South Korea

\item\label{InstOxford}
Mathematical Institute, University of Oxford\\
24--29 St.\ Giles', Oxford OX1 3LB, U.K.

\item\label{InstIAS}
School of Natural Sciences, Institute for Advanced Study,
Princeton, NJ 08540, USA

\item\label{InstBudapest}
Theoretical Physics Research Group of the
Hungarian Academy of Sciences\\
1117 P\'azm\'any s.\ 1/A, Budapest, Hungary

\item\label{InstCERN}
PH-TH Division, CERN, Geneva, Switzerland

\item\label{InstLAPTH}
LAPTH, Universit\'e de Savoie, CNRS\\
B.P.\ 110, 74941 Annecy-le-Vieux Cedex, France

\item\label{InstUppsala}
Department of Physics and Astronomy, 
Division for Theoretical Physics\\
Uppsala University, Box 516,
SE-751 08 Uppsala, Sweden

\item\label{InstKings}
Department of Mathematics, King's College, The Strand, London, WC2R 2LS, U.K.

\item\label{InstPNPI}
PNPI, Gatchina, Leningrad District, 188300 St.\ Petersburg, Russia

\item\label{InstKrakow}
Institute of Physics, Jagiellonian University\\
ul.\ Reymonta 4, 30-059 Krak\'ow, Poland

\item\label{InstENSP}
LPT, Ecole Normale Superi\'eure, 24 rue Lhomond, 75231 Paris Cedex 05, France

\item\label{InstParis6}
Universit\'e Pierre et Marie Curie (Paris-VI), 4 Place Jussieu, 75252 Paris Cedex 05, France

\item\label{InstPCTP}
Princeton Center for Theoretical Science\\
Princeton University, Princeton, NJ 08544, USA

\item\label{InstSaclay}
Institut de Physique Th\'eorique, CEA Saclay\\
91191 Gif-sur-Yvette Cedex, France

\item\label{InstNBI}
The Niels Bohr Institute, Blegdamsvej 17, 2100 Copenhagen, Denmark

\item\label{InstENSL}
Universit\'e de Lyon, Laboratoire de Physique, ENS Lyon et CNRS UMR 5672,\\
46 all\'ee d'Italie, 
69364 Lyon CEDEX 07, 
France

\item\label{InstFlorida}
Physics Department, P.O.~Box 248046, University of Miami\\
Coral Gables, FL 33124, USA

\item\label{InstIC}
Blackett Laboratory, Imperial College London, SW7 2AZ, U.K.

\item\label{InstPSU}
Department of Physics, The Pennsylvania State University\\
University Park, PA 16802, USA

\item\label{InstKITP}
Kavli Institute for Theoretical Physics,
University of California,\\
Santa Barbara,
CA 93106, USA

\item\label{InstHU}
Institut f\"ur Mathematik und Institut f\"ur Physik, Humboldt-Universit\"at zu Berlin\\
Johann von Neumann-Haus, Rudower Chaussee 25, 12489 Berlin, Germany

\item\label{InstNBIA}
Niels Bohr International Academy,
Niels Bohr Institute\\
Blegdamsvej 17,
2100 Copenhagen,
Denmark

\item\label{InstYork}
Department of Mathematics, University of York \\
Heslington, York, YO10 5DD, U.K.

\item\label{InstUtrecht}
Institute for Theoretical Physics, 
Utrecht University\\
Leuvenlaan 4,
3584 CE Utrecht,
The Netherlands

\item\label{InstPITP}
Perimeter Institute for Theoretical Physics,\\
Waterloo, Ontario N2L 2Y5, Canada

\end{list}
\endgroup

\newpage 


\setcounter{secnumdepth}{0}

\ifarxiv

\section{Preface}

\ifarxiv\else\begingroup\bfseries This is the preface for the
LMP special issue. It is not part of the present chapter.\par\endgroup\fi

Since late 2002 tremendous and rapid progress 
has been made in exploring 
planar $\superN=4$ super Yang--Mills theory 
and free IIB superstrings on the $AdS_5\times S^5$ background.
These two models are claimed to be exactly dual 
by the AdS/CFT correspondence, 
and the novel results give full support to the duality. 
The key to this progress lies in the 
integrability of the free/planar sector of the 
AdS/CFT pair of models.

Many reviews of integrability in the
context of the AdS/CFT correspondence are available in the literature.
They cover selected branches of the subject
which have appeared over the years.
Still it becomes increasingly difficult to maintain an overview 
of the entire subject, even for experts.
Already for several years there has been a clear demand for 
an up-to-date review to present a global view and summary of the subject,
its motivation, techniques, results and implications. 

Such a review appears to be a daunting task:
With around 8 years of development
and perhaps up to 1000 scientific articles written,
the preparation would represent a major burden on the prospective authors.
Therefore, our idea was to prepare a coordinated review collection to 
fill the gap of a missing global review for AdS/CFT integrability.
Coordination consisted in carefully splitting up the subject 
into a number of coherent topics.
These cover most aspects of the subject without overlapping too much.
Each topic is reviewed by someone 
who has made important contributions to it.
The collection is aimed at beginning students 
and at scientists working on different subjects, 
but also at experts who would like to (re)acquire an overview.
Special care was taken to keep the chapters brief (around 20 pages),
focused and self-contained
in order to enable the interested reader to 
absorb a selected topic in one go.

As the individual chapters will not convey
an overview of the subject as a whole, 
the purpose of the introductory chapter is 
to assemble the pieces of the puzzle into a bigger picture.
It consists of two parts: 
The first part is a general review of AdS/CFT integrability. 
It concentrates on setting the scene, outlining the achievements
and putting them into context.
It tries to provide a qualitative understanding
of what integrability is good for and how and why it works.
The second part maps out how the topics/chapters fit together
and make up the subject.
It also contains sketches of the contents of each chapter.
This part helps the reader in identifying the chapters
(s)he is most interested in.

There are reasons for and against combining all the contributions
into one article or book.
Practical issues however make it advisable
to have the chapters appear as autonomous review articles. 
After all, they are the works of individuals.
They are merely tied together by the introductory chapter
on which all the contributors have signed as coauthors.
If you wish to refer to this 
review on AdS/CFT integrability as a whole, 
we suggest that you cite (only) the introductory chapter:
\begin{center}\begin{minipage}{\linewidth}\centering
N.~Beisert et al.,\\
\textit{``Review of AdS/CFT Integrability: An Overview''},\\
\textsf{\doiref{10.1007/s11005-011-0529-2}{Lett.~Math.~Phys.~99,~3~(2012)}},
\texttt{\arxivref{1012.3982}{arXiv:1012.3982}}.
\end{minipage}\end{center}
If your work refers to a particular topic of the review, 
we encourage you to cite the 
corresponding specialised chapter(s) (instead/in addition), e.g.\
\begin{center}\begin{minipage}{\linewidth}\centering
J.~A.~Minahan,\\
\textit{``Review of AdS/CFT Integrability, Chapter I.1: Spin Chains in
  $\mathcal{N}$ = 4 SYM''},\\
\textsf{\doiref{10.1007/s11005-011-0522-9}{Lett.~Math.~Phys.~99,~33~(2012)}},
\texttt{\arxivref{1012.3983}{arXiv:1012.3983}}.
\end{minipage}\end{center}

Finally, I would like to thank my coauthors for their collaboration
on this project.
In particular, I am grateful to Pedro Vieira
who set up a website for internal discussions
which facilitated the coordination greatly:
Drafts and outlines of the chapters were uploaded to this forum.
Here, the contributors to the collection gave helpful 
comments and suggestions on the other chapters. 
It is fair to say that the forum 
improved the quality and completeness of the articles and how they fit together
before they first appeared in public. 
Also managing the final production stage would not have been 
nearly as efficient without it. Thanks for all your help 
and prompt availability during the last week!

\bigskip

\noindent
\qquad \textit{Niklas Beisert}\qquad Potsdam, December 2010




\newpage

\fi

\section{Introduction}
\label{sec:intro}


An old dream of Quantum Field Theory (QFT)
is to derive a quantitative description of 
the mass spectrum of hadronic particles and their excitations.
Ideally, one would be able to express the masses
of particles such as protons and neutrons
as functions of the parameters of the theory
\[
m\indup{p}=f_1(\alpha\indup{s},\alpha,\mu\indup{reg},\ldots),
\qquad
m\indup{n}=f_2(\alpha\indup{s},\alpha,\mu\indup{reg},\ldots),
\qquad
\ldots\,.
\]
They might be combinations of elementary functions, 
solutions to differential or integral equations
or something that can be evaluated effortlessly on a present-day computer.
For the energy levels of the hydrogen atom
analogous functions are known and they work to a high accuracy.
However, it has become clear that an elementary analytical understanding 
of the hadron spectrum will remain a dream.
There are many reasons why this is more than can be expected;
just to mention a few:
At low energies,
the coupling constant $\alpha\indup{s}$ is too large 
for meaningful approximations.
In particular, non-perturbative contributions dominate
such that the standard loop expansion 
simply does not apply.
Self-interactions of the chromodynamic field 
lead to a non-linear and highly complex problem.
Clearly, confinement obscures the nature of 
fundamental particles in Quantum Chromodynamics (QCD) at low energies.
Of course there are non-perturbative methods 
to arrive at reasonable approximations for the spectrum, 
but these are typically based on effective field theory 
or elaborate numerical simulations instead of elementary analytical QCD.

\paragraph{Spectrum of Scaling Dimensions.}

We shall use the above hadronic spectrum as an analog 
to explain the progress in applying methods of integrability
to the spectrum of planar $\superN=4$ super Yang--Mills (SYM) theory.%
\footnote{Please note that, here and below, references to the original literature 
can be found in the chapters of this review collection
where the underlying models are introduced.}
The analogy does not go all the way, 
certainly not at a technical level,
but it is still useful for a qualitative understanding
of the achievements.

First of all, $\superN=4$ SYM 
is a cousin of QCD and of the Standard Model
of particle physics. 
It is based on the same types of fundamental particles and interactions
--- it is a renormalisable gauge field theory 
on four-dimensional Minkowski space --- 
but the details of the models are different.
Importantly, $\superN=4$ SYM has a much richer set of symmetries:
supersymmetry and conformal symmetry. 
In particular, the latter implies that there are no massive particles
whose spectrum we might wish to compute. 
Nevertheless, composite particles and their mass spectrum have
an analogue in conformal field theories: 
These are called \emph{local operators}.
They are composed from the fundamental fields,
all residing at a common point in spacetime. 
As in QCD, the colour charges are balanced out 
making the composites gauge-invariant objects.
Last but not least, there is a characteristic quantity
to replace the mass, the so-called \emph{scaling dimension}.
Classically, it equals the sum of the constituent dimensions,
and, like the mass, it does receive quantum corrections
(the so-called \emph{anomalous} dimensions)
from interactions between the constituents.

In the planar $\superN=4$ SYM model and for scaling dimensions of local operators
the particle physicist's dream is coming true.
We know how to express the scaling dimension $D_O$ of some local operator $O$
as a function of the coupling constant $\lambda$ 
\[
D_O=f(\lambda).
\]
In general this function is given as the solution of a
set of integral equations.%
\footnote{As a matter of fact, the system of equations
is not yet in a form which enables easy evaluation. 
E.g.~there are infinitely many equations for infinitely many quantities.
It is however common belief that one can, as in similar cases, 
reduce the system to a finite set of Non-Linear Integral Equations (NLIE).}
What is more, in particular cases the equations 
have been solved numerically for a wide range of $\lambda$'s!
These equations follow from the so-called Thermodynamic Bethe Ansatz (TBA)
or related techniques (Y-system).
In a certain limit, the equations simplify
to a set of algebraic equations,
the so-called asymptotic Bethe equations.
It is also becoming clear that not only the spectrum, 
but many other observables can be determined in this way.
Thus it appears that planar $\superN=4$ SYM can be \emph{solved} exactly.

\paragraph{Integrability.}

With the new methods at hand we can now compute observables
which were previously inaccessible by all practical means.
By studying the observables and the solution, 
we hope to get novel insights, not only into this particular model, 
but also into quantum gauge field theory in general. 
What is it that makes planar $\superN=4$ SYM calculable 
and other models not? 
Is its behaviour generic or very special?
Can we for instance use the solution 
as a starting point or first approximation for other models? 
On the one hand one may view $\superN=4$ SYM as a very special QFT. 
On the other hand, any other four-dimensional gauge theory can be viewed
as $\superN=4$ SYM with some particles and interactions added or removed:
For instance, several quantities show a universal behaviour 
throughout the class of four-dimensional gauge theories
(e.g.\ highest ``transcendentality'' part, tree-level gluon scattering).
Moreover this behaviour is dictated by $\superN=4$ SYM 
acting as a representative model.
Thus, indeed, selected results obtained in $\superN=4$ SYM 
can be carried over to general gauge theories.
Nevertheless it is obvious that we cannot make direct predictions
along these lines for most observables, such as the hadron spectrum.

The miracle which leads to the solution of planar $\superN=4$ SYM 
described above is generally called \emph{integrability}.
Integrability is a phenomenon which is typically confined to 
two-dimensional models (of Euclidean or Minkowski signature).
Oddly, here it helps in solving a four-dimensional QFT.
%

\paragraph{AdS/CFT Correspondence.}

A more intuitive understanding of why there is integrability 
comes from the \emph{AdS/CFT correspondence}
\cite{Maldacena:1998re,Gubser:1998bc,Witten:1998qj},
see also the reviews \cite{Kovacs:1999fx,Aharony:1999ti,D'Hoker:2002aw,Maldacena:2003nj,Nastase:2007kj,Polchinski:2010hw}
and \cite{Benna:2008yg,Klebanov:2009zz}.
The latter is a duality relation between certain pairs of models.
One partner is a conformal field theory, 
i.e.\ a QFT with exact conformal spacetime symmetry.
The other partner is a string theory 
where the strings propagate on a background which 
contains an Anti-de-Sitter spacetime (AdS) as a factor.
The boundary of an $AdS_{d+1}$ spacetime 
is a conformally flat $d$-dimensional spacetime
on which the CFT is formulated.
The AdS/CFT duality relates the string partition function 
with sources $\phi$ for string vertex operators 
fixed to value $J$ at the boundary of $AdS_{d+1}$ 
to the CFT$_d$ partition function with sources $J$ for local operators 
\[
Z\indup{str}[\phi|_{\partial AdS}=J] = Z\indup{CFT}[J].
\]
More colloquially: 
For every string observable at the boundary 
of $AdS_{d+1}$ there is a corresponding 
observable in the CFT$_d$ (and vice versa)
whose values are expected to match.
This is a remarkable statement
because it relates two rather different types 
of models on spacetimes of different dimensionalities.
From it we gain novel insights into one model 
through established results from the other model.
For example, we can hope to learn about 
the long-standing problem of quantum gravity 
(gravity being a fundamental part of every string theory)
through studying a more conventional QFT. 
However, this transfer of results 
requires a leap of faith
as long as the duality lacks a formal proof. 

Most attempts at testing the predictions of the AdS/CFT duality
have focused on its most symmetric setting:
The CFT partner is the gauge theory featured above, $\superN=4$ SYM.
The string partner is IIB superstring theory on the $AdS_5\times S^5$ background.
This pair is an ideal testing ground because the large amount of supersymmetry 
leads to simplifications 
and even allows for exact statements about both models.
In this context, we can also understand 
the miraculous appearance of integrability in planar $\superN=4$ SYM better:
By means of the AdS/CFT duality it translates 
to integrability of the string worldsheet model.
The latter is a two-dimensional non-linear sigma model 
on a symmetric coset space
for which integrability is a common phenomenon.
Consequently, integrability has become an important tool to 
perform exact calculations in both models.
Full agreement between both sides of the duality 
has been observed in all considered cases.
Therefore, integrability has added substantially 
to the credibility of the AdS/CFT correspondence.

\paragraph{String/Gauge Duality.}

Another important aspect of the AdS/CFT duality is that
in many cases it relates a string theory to a gauge theory.
In fact, the insight regarding the similarities between these two types of models
is as old as string theory:
It is well-known that the hadron spectrum organises
into so-called Regge trajectories.
These represent an approximate linear relationship with universal slope 
between the mass squared of hadronic resonances and their spin. 
This is precisely what a string theory on flat space predicts,
hence string theory was for some time considered a candidate model
of the strong interactions. 
For various reasons this idea did not work out.
Instead, it was found that a gauge theory, 
namely QCD, provides an accurate and self-consistent 
description of the strong interactions.
Altogether it implies that string theory, under some conditions,
can be a useful approximation to gauge theory phenomena.
A manifestation of stringy behaviour in gauge theory 
is the occurrence of flux tubes of the chromodynamic field.
Flux tubes form between two quarks when they are pulled apart.
To some approximation they can be viewed as one-dimensional objects 
with constant tension, i.e.\ strings.
The AdS/CFT correspondence goes even further. 
It proposes that in some cases a gauge theory is exactly
dual to a string theory. By studying those cases, 
we hope to gain more insights into string/gauge duality in general,
perhaps even for QCD.

\begin{figure}
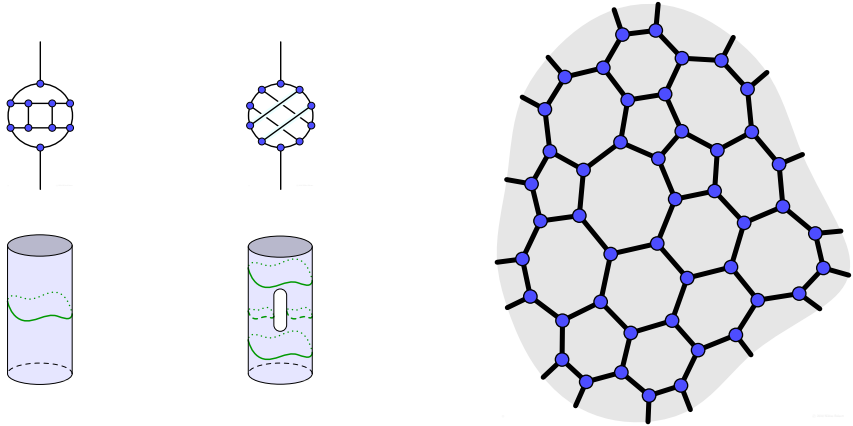
\centering
\begin{minipage}{0.4\linewidth}
\parbox{0.5\linewidth}{\centering\includegraphics[height=2cm]{FigPlanarG0.mps}}%
\parbox{0.5\linewidth}{\centering\includegraphics[height=2cm]{FigPlanarG1.mps}}%
\par\vspace{3ex}
\parbox{0.5\linewidth}{\centering\includegraphics[height=2cm]{FigPlanarS0.mps}}%
\parbox{0.5\linewidth}{\centering\includegraphics[height=2cm]{FigPlanarS1.mps}}%
\end{minipage}
\hspace{0.06\textwidth}
\includegraphicsbox[width=0.3\textwidth]{FigPlanarPlane.mps}
\caption{Planar and non-planar Feynman graph (top),
free and interacting string worldsheet (bottom), 
Feynman graph corresponding to a patch of worldsheet (right).}
\label{fig:PlanarLimit}
\end{figure}

A milestone of string/gauge duality 
was the discovery of the planar limit
\cite{'tHooft:1973jz}, see \figref{fig:PlanarLimit}.
This is a limit for models with gauge group
$\grp{SU}(N\indup{c})$, $\grp{SO}(N\indup{c})$ or $\grp{Sp}(N\indup{c})$. 
It consists in taking the rank of the group to infinity,
$N\indup{c}\to\infty$,
while keeping the rescaled gauge coupling $\lambda=g\indup{YM}^2N\indup{c}$ finite.
In this limit, the Feynman graphs which describe the perturbative expansion 
of gauge theory around $\lambda=0$ can be classified according to their genus:
Graphs which can be drawn on the plane without crossing lines are called planar.
The remaining graphs with crossing lines are suppressed.
This substantially reduces the complexity of graphs
from factorial to exponential growth,
such that the radius of convergence 
of the perturbative series grows 
to a finite size.
Moreover, the surface on which the Feynman graphs are drawn
introduces a two-dimensional structure into gauge theory:
It is analogous to the worldsheet of a string
whose string coupling $g\indup{str}$ is proportional to $1/N\indup{c}$.
Not surprisingly, integrability is confined to this planar limit
where gauge theory resembles string theory.

\paragraph{Parameter Space.}

\begin{figure}\centering
\includegraphics[width=0.7\textwidth]{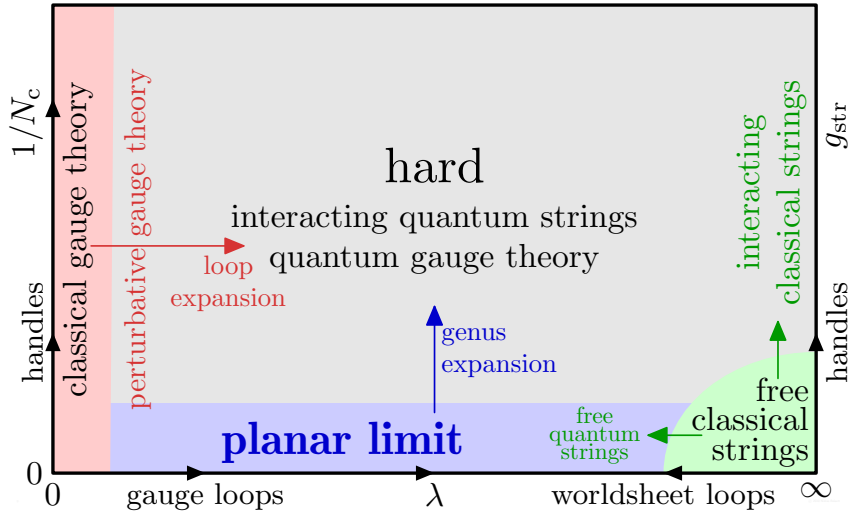}
\caption{Map of the parameter space of
$\superN=4$ SYM or strings on $AdS_5\times S^5$.}
\label{fig:ParameterSpace}
\end{figure}

Let us now discuss the progress due to integrability 
based on a map of the parameter space 
of our gauge and string theory, see \figref{fig:ParameterSpace}.
Typically there are two relevant parameters for a gauge theory, 
the 't Hooft coupling $\lambda=g\indup{YM}^2N\indup{c}$
and the number of colours $N\indup{c}$ as a measure of the rank of the gauge group.
In a string theory we have the effective string tension $T=R^2/2\pi\alpha'$
(composed from the inverse string tension $\alpha'$ and the $AdS_5$/$S^5$ radius $R$)
and the string coupling $g\indup{str}$.
The AdS/CFT correspondence relates them as follows
\[
\lambda=4\pi^2T^2\,,
\qquad
\frac{1}{N\indup{c}}=
\frac{g\indup{str}}{4\pi^2T^2}\,.
\]
The region of parameter space where $\lambda$ is small 
is generally called the \emph{weak coupling} regime.
This is where perturbative gauge theory 
in terms of Feynman diagrams provides reliable results. 
By adding more loop orders to the series expansion
one can obtain more accurate estimates towards the centre
of the parameter space (up to non-perturbative effects).
Unfortunately, conventional methods in combination with computer algebra
only allow evaluating the first few coefficients of the series in practice.
Thus we cannot probe the parameter space far away from the weak coupling regime.
However, $N\indup{c}$ can be finite in practice, therefore the regime of
perturbative gauge theory extends along the line $\lambda=0$.

The region around the point $\lambda=\infty$, $g\indup{str}=0$
is where perturbative string theory applies.
Here, strings are weakly coupled, 
but the region is nevertheless called the \emph{strong coupling} regime 
referring to the gauge theory parameter $\lambda$.
String theory provides a double expansion around this point. 
The accuracy towards finite $\lambda$ is increased by 
adding quantum corrections to the worldsheet sigma model
(curvature expansion, ``worldsheet loops'').
Finite-$g\indup{str}$ corrections correspond to 
adding handles to the string worldsheet
(genus expansion, ``string loops'').
As before, both expansions are far from trivial, and typically
only the first few coefficients can be computed in practice.
Consequently, series expansions do not give reliable approximations
far away from the point $\lambda=\infty$, $g\indup{str}=0$.

\begin{figure}
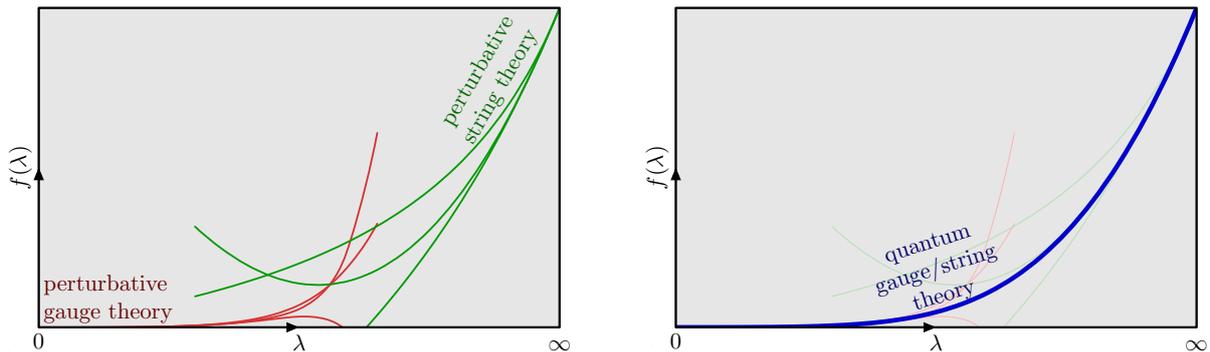
\centering
\includegraphics[width=0.47\textwidth]{FigInterpolateE.mps}%
\hfill
\includegraphics[width=0.47\textwidth]{FigInterpolateI.mps}%
\caption{Weak coupling (3, 5, 7 loops) and strong coupling (0, 1, 2 loops) expansions
(left) and numerically exact evaluation (right)
of some interpolating function $f(\lambda)$.}
\label{fig:Interpolate}
\end{figure}

Here we can see the weak/strong dilemma of the AdS/CFT duality,
see also \figref{fig:Interpolate}:
The perturbative regimes of the two models do not overlap.
On the one hand AdS/CFT provides novel insights into both models. 
On the other hand, we cannot really be sure of them 
until there is a general proof of the duality.
Conventional perturbative expansions are of limited use in verifying,
and tests had been possible only for a few special observables
(cf.\ \cite{Horowitz:2006ct} for example).

This is where integrability comes to help.
As explained above, it provides novel computational means 
in \emph{planar} $\superN=4$ SYM at arbitrary coupling $\lambda$. 
The AdS/CFT correspondence relates this regime to 
free ($g\indup{str}=0$) IIB superstrings 
on $AdS_5\times S^5$ at arbitrary tension $T$.
It connects the regime of perturbative gauge theory 
with the regime of perturbative string theory.
Integrability predicts the spectrum of planar scaling dimensions
for local operators as a function of $\lambda$, cf.\ \figref{fig:Interpolate}.
In string theory this is dual to the energy spectrum 
of free string states 
(strings which neither break apart nor join with others). 
We find that integrability makes coincident predictions for both models.
At weak coupling one can compare to results
obtained by conventional perturbative means in gauge theory,
and one finds agreement.
Analogous agreement with perturbative strings is found at strong coupling.
And for intermediate coupling the spectrum apparently interpolates smoothly
between the two perturbative regimes.

Methods of integrability provide us with reliable data
over the complete range of couplings. 
We can investigate in practice a gauge theory 
at strong coupling. There it behaves like a weakly coupled string theory. 
Likewise a string theory on a highly curved background 
(equivalent to low tension) behaves like a weakly coupled gauge theory.
At intermediate coupling, the results
are reminiscent of neither model or of both;
this is merely a matter of taste 
and depends crucially on whether 
one's intuition is based on classical or quantum physics.
In any case, integrability can give us valuable insights
into a truly quantum gauge and/or string theory
at intermediate coupling strength.

\paragraph{Solving a Theory.}

In conclusion, we claim that integrability solves 
the planar sector of a particular pair of gauge and string theories.
We should be clear about the actual meaning of this statement:
It certainly does not mean that the spectrum is given by
a simple formula as in the case of a harmonic oscillator,
the (idealised) hydrogen atom or strings in flat space
(essentially a collection of harmonic oscillators)%
\footnote{In fact, these systems are also integrable, 
but of an even simpler kind.}
\[
E\indup{HO}=\omega(n+\half),
\qquad
E\indup{hyd}=-\frac{m\indup{e}\alpha^2}{n^2}\,,
\qquad
m^2\indup{flat}=m^2_0+\frac{1}{\alpha'}\sum_{k=-\infty}^\infty n_k |k|.
\]
It would be too much to hope for such a simplistic behaviour in our models:
For instance, the one-loop corrections to scaling dimensions
are typically algebraic numbers. Therefore the best we can expect 
is to find a system of algebraic equations 
whose solutions determine the spectrum. 
This is what methods of integrability provide
more or less directly.
Integrability vastly reduces the complexity of the spectral problem
by bypassing almost all steps of standard QFT methods:
There we first need to compute
all the entries of the matrix of scaling dimensions. 
Each entry requires a full-fledged computation 
of higher loop Feynman graphs 
involving sophisticated combinatorics and demanding loop integrals.
The naively evaluated matrix contains infinities calling for proper regularisation and renormalisation.
The final step consists in diagonalising this (potentially large) matrix.
This is why scaling dimensions are solutions of algebraic equations.
In comparison, the integrable approach directly predicts the algebraic equations determining 
the scaling dimension $D$
\[
f(D,\lambda)=0.
\]
This is what we call a \emph{solution} of the spectral problem.

A crucial benefit of integrability is that the spectral equations 
include the coupling constant $\lambda$ in functional form. 
Whereas standard methods produce an expansion 
whose higher loop coefficients are exponentially or even factorially hard to compute,
here we can directly work at intermediate coupling strength. 

\begin{figure}\centering
\includegraphics[width=0.7\textwidth]{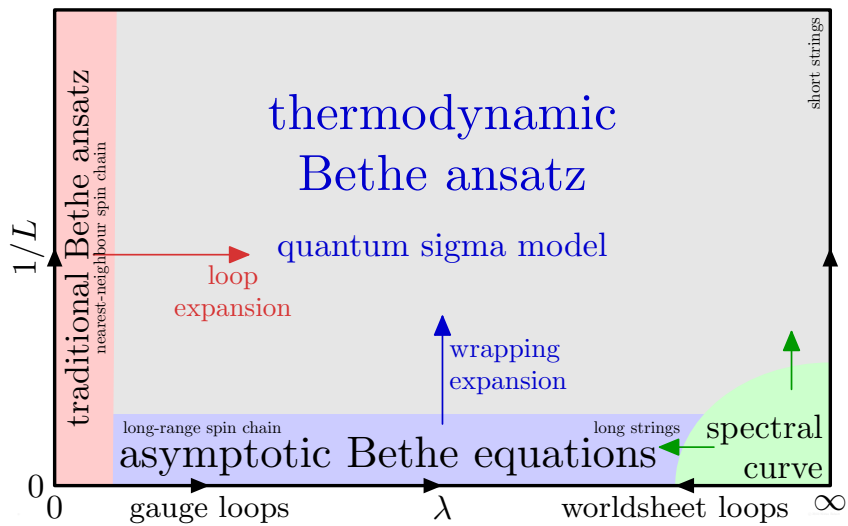}
\caption{Phase diagram of local operators in 
planar $\superN=4$ SYM mapped with respect to coupling $\lambda$ vs.\ ``size'' $L$.
Also indicated are the integrability methods 
that describe the spectrum accurately.}
\label{fig:Thermo}
\end{figure}

What is more, integrability gives us easy access to 
composite objects with a large number of constituents. 
Generally, there is an enormous phase space 
for such objects growing exponentially with their size.
Standard methods would require computing the complete 
matrix of scaling dimensions and then filter out the desired eigenvalue.
Clearly this procedure is prohibitive for large sizes.
Conversely, the integrable approach 
is formulated in terms of physically meaningful quantities. 
This allows us to assume a certain coherent behaviour for 
the constituents of the object we are interested in,
and then approach the \emph{thermodynamic limit}. 
Consequently we obtain a set of equations 
for the energy of just this object.
Moreover, the thermodynamic limit is typically much simpler
than the finite-size equations. 
The size of the object can be viewed as a quantum parameter,
where infinite vs.\ finite size corresponds to 
classical vs.\ quantum physics.
In fact, in many cases it does map to classical vs.\ quantum strings!
In \figref{fig:Thermo} we present a phase space 
for local operators in planar $\superN=4$ SYM. 
On it we indicate the respective integrability 
methods to be described in detail 
in the overview section and in the chapters.

As already mentioned, a strength of the integrable 
system approach is that objects are often
represented through their physical parameters.
This is not just an appealing feature, 
but also a reason for the efficiency:
The framework of quantum mechanics and QFT 
is heavily based on equivalence classes. 
Explicit calculations usually work with representatives.
Choosing a particular representative in a class introduces
further auxiliary degrees of freedom into the system. 
These degrees of freedom are carried along the intermediate steps of the calculation,
and it is reasonable to expect them to be a source of added difficulty
because there is no physical principle to constrain their contributions. 
In particular, they are the habitat of the notorious infinities of QFT.
At the end of the day, all of their contributions miraculously%
\footnote{Of course, the miracle is consistency of the model 
paired with failing to make mistakes in the calculation
(often used as a convenient cross check of the result).}
vanish into thin air. 
Hence a substantial amount of efforts
typically go into calculating contributions 
which one is actually not interested in.
Conversely, one may view integrable methods 
as working directly in terms of the physical equivalence classes 
instead of their representatives. 
The observables are then computed without intermediate steps 
or complications. 
The fact that such a shortcut exists for some models is a true miracle;
it is called \emph{integrability}.

So far we have discussed solving the spectrum of our planar model(s). 
A large amount of evidence has now accumulated 
that this is indeed possible, and, more importantly, 
we understand how to do it in practice.
Solving the theory, however, requires much more;
we should be able to compute \emph{all of its observables}.
For a gauge theory they include not only the spectrum
of scaling dimensions, but also correlation functions, 
scattering amplitudes, expectation values of Wilson loops,
surface operators and other extended objects,
as well as combinations of these (loops with insertions,
form factors, \ldots), if not more.
For several of these, in particular for scattering amplitudes,
it is becoming clear that integrability 
provides tools to substantially simplify their computation.
Hence it is plausible to expect that the planar limit can be solved. 

Can we also solve the models away from the planar limit?
There are many indications that integrability 
breaks down for finite number of colours $N\indup{c}$. 
Nevertheless, this alone does not imply that
we should become dispirited.
Integrability may still prove useful, 
not in the sense of an exact solution, 
but as a means to perform an expansion 
in terms of genus, i.e.\ in powers of $1/N\indup{c}\sim g\indup{str}$.
This might give us a new handle to approach the centre
of parameter space in \figref{fig:ParameterSpace} coming from below. 
The centre will, with all due optimism, most likely remain a tough nut to crack.

In conclusion, methods of integrability have already brought
and will continue to bring
novel insights into the gauge and string models.
Having many concrete results at hand 
helps in particular to understand their duality better. 
In particular we can confirm and complete the AdS/CFT \emph{dictionary}
which relates objects and observables between the two models.

\paragraph{Integrability as a Symmetry.}

Above we have argued that the success of integrability 
is based on the strict reduction to the physical degrees of freedom. 
Another important point of view is that integrability 
is a hidden symmetry. Symmetries have always been a key towards
a better understanding in particle physics and QFT.
Here the hidden symmetry is in fact so powerful that it
not only relates selected quantities to others, 
but, in some sense, anything to everything else. 
The extended symmetry thus predicts the outcome of every measurement,
at least in principle. 
Conventionally one would expect the resulting model to be trivial, 
just like a harmonic oscillator, 
but there are important interesting and highly non-trivial cases. 

Integrability finds a natural mathematical implementation 
in the field of \emph{quantum algebra}. 
More concretely, the type of quantum integrable system
that we encounter is usually formulated in terms of 
deformed universal enveloping algebras of affine Lie algebras.
The theory of such quasi-triangular Hopf algebras
is in general highly developed. 
It provides the objects and their relations for the solution 
of the physical system. 
Curiously, our gauge/string theory integrable model appears to 
be based on some unconventional or exceptional superalgebra which largely remains to be understood. 
It is not even clear whether quasi-triangular Hopf algebras 
are a sufficient framework for a complete mathematical implementation 
of the system.

\paragraph{Relations to Other Subjects.}

An aspect which makes the topic of this review 
a particularly attractive one to work on 
is its relation to diverse subjects of
theoretical physics and mathematics.
Let us collect a few here, including those mentioned above,
together with references to the chapters of this review 
where the relations are discussed in more detail:
\begin{itemize}
\item
Most obviously, the topic of the review itself belongs to 
\emph{four-dimensional QFT}, more specifically, \emph{gauge theory} and/or \emph{CFT},
but also to \emph{string theory} on \emph{curved backgrounds}. 

\item
Recalling the discussion from a few lines above, 
the mathematical framework for the kind of integrable models 
that we encounter is \emph{quantum algebra},
see \namedref{Chapter}{chap:Yang}.

\item
As mentioned earlier, string theory always
contains a self-consistent formulation of quantum gravity.
By gaining a deeper understanding of string theory models, 
we hope to learn more about \emph{quantum gravity} as such.
Furthermore, by means of the string-related Kawai--Lewellen--Tye \cite{Kawai:1985xq}
and Bern--Carrasco--Johansson relations \cite{Bern:2008qj,Bern:2010ue},
there is a connection between scattering amplitudes 
in $\superN=4$ SYM and \emph{$\superN=8$ supergravity},
which stands a chance of being free of perturbative divergencies.%
\footnote{One should point out that these relations 
are essentially non-planar.}
These aspects are not part of the review. In fact, it would be highly desirable
to explore the use of integrability results in this context.

\item
Prior to the discoveries related to the AdS/CFT correspondence,
integrability in four-dimensional gauge theories was already observed 
in the context of \emph{high-energy scattering} and the BFKL equations,
and for \emph{deep inelastic scattering} and the DGLAP equations,
see \namedref{Chapter}{chap:QCD} and \cite{Ioffe:2010zz}.
Note that the \emph{twist states}
discussed in \namedref{Chapters}{chap:Twist} 
play a prominent role in deep inelastic scattering.

\item
There are also rather distinct applications of integrability 
in supersymmetric gauge theories: 
There is the famous Seiberg--Witten solution \cite{Seiberg:1994rs,Seiberg:1994aj}
for the BPS masses in $D=4$, $\superN=2$ gauge theories.
Furthermore, supersymmetric vacua in $D=2$, $\superN=4$ gauge theories with matter
can be described by Bethe ans\"atze \cite{Nekrasov:2009uh,Nekrasov:2009ui}.
It remains to be seen whether there are connections
to the subject of the present review.

\item
There are further links to 
general \emph{four-dimensional gauge theories}:
On a qualitative level we might hope to learn about QCD strings
from novel results in the AdS/CFT correspondence at finite coupling.
On a practical level, the leading-order results in $\superN=4$ SYM 
can be carried over to general gauge theories
essentially because $\superN=4$ SYM contains all types of particles 
and interactions allowed in a renormalisable QFT. 
\namedref{Chapter}{chap:QCD} is most closely related to this topic.

\item
Along the same lines, the BFKL dynamics in leading logarithmic approximation
is universal to all four-dimensional gauge theories. 
The analytic expressions derived in $\superN=4$ SYM may allow us to clarify 
the nature of the most interesting Regge singularity, 
\emph{the pomeron} (see \cite{Lotter:1996vk,Forshaw:1997dc,Donnachie:2002en}), 
which is the most interesting object for perturbative QCD 
and for its applications to particle collider physics.%
\footnote{We thank L.~Lipatov for pointing out this application.}

\item
A certain class of composite states, 
but also loop integrals in QFT, often involve 
generalised harmonic sums, generalised polylogarithms and multiple zeta values.
The exploration of such \emph{special functions} is an active topic of mathematics.
See \namedref{Chapters}{chap:Higher}, \ref{chap:Twist} \namedref{and}{chap:Dual}.

\item
Local operators of the gauge theory are equivalent to
states of a quantum spin chain. 
Spin chain models come to use 
in connection with \emph{magnetic properties} in \emph{solid state physics}. 
Also in gauge theory, 
ferromagnetic and anti-ferromagnetic states 
play an important role,
see \namedref{Chapters}{chap:Chain}
\namedref{and}{chap:Twist}.

\item
More elaborate spin chains
--- such as the one-dimensional Hubbard model 
(cf.\ \cite{Essler:2005:aa}) ---
are considered in connection to \emph{electron transport}. 
Curiously, this rather exceptional Hubbard chain 
makes an appearance in the gauge theory context,
in at least two distinct ways, 
see \namedref{Chapters}{chap:LR} \namedref{and}{chap:SMat}.


\item
And there are many more avenues left to be explored.

\end{itemize}

\newpage

\section{Outline}

The review collection consists of the above \hypref{sec:intro}{introduction}
and \ifarxiv\hypref{sec:toc}{23 chapters}\else 23 chapters\fi\ grouped into 6 major subjects.
Each chapter reviews a particular topic in a self-contained manner.
The following \hypref{sec:overview}{overview} 
gives a brief summary of each part and each chapter,
and is meant to tie the whole collection together. 
It can be understood as an extensive table of contents.

Where possible, we have put the chapters into a natural and 
meaningful order with regards to content. 
A chapter builds upon insights and results presented in the earlier chapters
and begins roughly where the previous one ended.
In many cases this reflects the historical developments, 
but we have tried to pull loops straight. 
Our aim was to prepare a pedagogical and generally accessible introduction
to the subject of AdS/CFT integrability rather than a historically accurate account.

While the topics were fixed, 
the design and presentation of each chapter 
was largely the responsibility of its authors.
The only guideline was to discuss 
an instructive example in detail
while presenting the majority of results more briefly.
Furthermore, the chapters give a guide to the literature
relevant to the topic where more details can be found.
Open problems are also discussed in the chapters.
Note that we did not enforce uniform conventions
for naming, use of $\mathfrak{alphabets}$, normalisations, and so on.
This merely reflects a reality of the literature.
However, each chapter is meant to be self-consistent. 

Before we begin with the overview, we would like to point out 
existing reviews on AdS/CFT integrability and related subjects
which cover specific aspects in more detail.
We can recommend several reviews dedicated to the subject
\cite{Plefka:2005bk,Minahan:2006sk,Dorey:2009zz,Arutyunov:2009ga,Basso:2009gh,Alday:2009zza,Serban:2010sr,Fiamberti:2010fw}.
Also a number of PhD theses are available 
which at least contain a general review as the introduction
\cite{Beisert:2004ry,Swanson:2007dh,Okamura:2008jm,Vicedo:2008jk,Rej:2009je,Gromov:2009zz,Volin:2010cq,Puletti:2010ge,deLeeuw:2010nd}.
It is also worthwhile to read 
some of the very brief accounts of the subject
in the form of news items \cite{SchaferNameki:2006ky,Nicolai:2007zza}.
Last but not least, we would like to refer the reader to prefaces 
of special issues dedicated to AdS/CFT integrability \cite{Tseytlin:2009zz,Tseytlin:2011zz}
and closely related subjects \cite{Dorey:2006zz,Alcaraz:2006zz}.

\newpage

\ifarxiv

\section{Table of Contents}
\label{sec:toc}

page numbers refer to \textsf{Lett.~Math.~Phys.~99~(2012)},
numbers 1210.0392--0405 are \href{http://arxiv.org}{arxiv.org} identifiers.

\begin{list}{}%
{\settowidth\labelwidth{III.1}%
 \leftmargin\labelwidth%
 \advance\leftmargin\labelsep%
 \parsep 0pt
 \raggedright
 \small
 \setcounter{enumi}{0}%
 \usecounter{enumii}%
 \renewcommand{\theenumi}{\Roman{enumi}}%
 \renewcommand{\theenumii}{.\arabic{enumii}}%
 \newcommand{\partitem}[1]{%
  \vspace{1ex}%
  \stepcounter{enumi}%
  \item[\textbf{\hypref{#1}{\theenumi}}\hfill]%
  \setcounter{enumii}{0}%
 }
 \newcommand{\chapitem}[1]{%
  \refstepcounter{enumii}%
  \item[\hypref{#1}{\theenumi\theenumii}\hfill]%
 }
}

\partitem{part:gauge}
\textbf{\hypref{part:gauge}{$\superN=4$ Super Yang--Mills Theory}}

\chapitem{chap:Chain}\label{chapChain}
J.~A.~Minahan,
\textit{``\hypref{chap:Chain}{Spin Chains in $\mathcal{N}$ = 4 SYM}''},
\doiref{10.1007/s11005-011-0522-9}{\textsf{pp.~33}}, 
\texttt{\arxivref{1012.3983}{1012.3983}}.

\chapitem{chap:Higher}\label{chapHigher}
C.~Sieg,
\textit{``\hypref{chap:Higher}{The spectrum from perturbative gauge theory}''},
\doiref{10.1007/s11005-011-0508-7}{\textsf{pp.~59}},
\texttt{\arxivref{1012.3984}{1012.3984}}.

\chapitem{chap:LR}\label{chapLR}
A.~Rej,
\textit{``\hypref{chap:LR}{Long-range spin chains}''},
\doiref{10.1007/s11005-011-0509-6}{\textsf{pp.~85}},
\texttt{\arxivref{1012.3985}{1012.3985}}.

\partitem{part:string} 
\textbf{\hypref{part:string}{IIB Superstrings on $AdS_5\times S^5$}}

\chapitem{chap:Spinning}\label{chapSpinning}
A.~Tseytlin,
\textit{``\hypref{chap:Spinning}{Classical $AdS_5\times S^5$ string solutions}''},
\doiref{10.1007/s11005-011-0466-0}{\textsf{pp.~103}},
\texttt{\arxivref{1012.3986}{1012.3986}}.

\chapitem{chap:QString}\label{chapQString}
T.~McLoughlin,
\textit{``\hypref{chap:QString}{Quantum Strings in $AdS_5\times S^5$}''},
\doiref{10.1007/s11005-011-0510-0}{\textsf{pp.~127}},
\texttt{\arxivref{1012.3987}{1012.3987}}.

\chapitem{chap:Sigma}\label{chapSigma}
M.~Magro,
\textit{``\hypref{chap:Sigma}{Sigma Model, Gauge Fixing}''},
\doiref{10.1007/s11005-011-0481-1}{\textsf{pp.~149}},
\texttt{\arxivref{1012.3988}{1012.3988}}.

\chapitem{chap:Curve}\label{chapCurve}
S.~Sch{\"a}fer-Nameki,
\textit{``\hypref{chap:Curve}{The Spectral Curve}''},
\doiref{10.1007/s11005-011-0525-6}{\textsf{pp.~169}},
\texttt{\arxivref{1012.3989}{1012.3989}}.

\partitem{part:spectrum}
\textbf{\hypref{part:spectrum}{Solving the AdS/CFT Spectrum}}

\chapitem{chap:ABA}\label{chapABA}
M.~Staudacher,
\textit{``\hypref{chap:ABA}{Bethe Ans\"atze and the R-Matrix Formalism}''},
\doiref{10.1007/s11005-011-0530-9}{\textsf{pp.~191}},
\texttt{\arxivref{1012.3990}{1012.3990}}.

\chapitem{chap:SMat}\label{chapSMat}
C.~Ahn and R.~I.~Nepomechie,
\textit{``\hypref{chap:SMat}{Exact world-sheet S-matrix}''},
\doiref{10.1007/s11005-011-0478-9}{\textsf{pp.~209}},
\texttt{\arxivref{1012.3991}{1012.3991}}.

\chapitem{chap:SProp}\label{chapSProp}
P.~Vieira and D.~Volin,
\textit{``\hypref{chap:SProp}{The dressing factor}''},
\doiref{10.1007/s11005-011-0482-0}{\textsf{pp.~231}},
\texttt{\arxivref{1012.3992}{1012.3992}}.

\chapitem{chap:Twist}\label{chapTwist}
L.~Freyhult,
\textit{``\hypref{chap:Twist}{Twist states and the cusp anomalous dimension}''},
\doiref{10.1007/s11005-011-0483-z}{\textsf{pp.~255}},
\texttt{\arxivref{1012.3993}{1012.3993}}.

\chapitem{chap:Luescher}\label{chapLuescher}
R.~Janik,
\textit{``\hypref{chap:Luescher}{L\"uscher corrections}''},
\doiref{10.1007/s11005-011-0511-z}{\textsf{pp.~277}},
\texttt{\arxivref{1012.3994}{1012.3994}}.

\chapitem{chap:TBA}\label{chapTBA}
Z.~Bajnok,
\textit{``\hypref{chap:TBA}{Thermodynamic Bethe Ansatz}''},
\doiref{10.1007/s11005-011-0512-y}{\textsf{pp.~299}},
\texttt{\arxivref{1012.3995}{1012.3995}}.

\chapitem{chap:Trans}\label{chapTrans}
V.~Kazakov and N.~Gromov,
\textit{``\hypref{chap:Trans}{Hirota Dynamics for Quantum Integrability}''},
\doiref{10.1007/s11005-011-0513-x}{\textsf{pp.~321}},
\texttt{\arxivref{1012.3996}{1012.3996}}.

\partitem{part:beyond}
\textbf{\hypref{part:beyond}{Further Applications of Integrability}}

\chapitem{chap:Observ}\label{chapObserv}
C.~Kristjansen,
\textit{``\hypref{chap:Observ}{Aspects of Non-planarity}''},
\doiref{10.1007/s11005-011-0514-9}{\textsf{pp.~349}},
\texttt{\arxivref{1012.3997}{1012.3997}}.

\chapitem{chap:Deform}\label{chapDeform}
K.~Zoubos,
\textit{``\hypref{chap:Deform}{Deformations, Orbifolds and Open Boundaries}''},
\doiref{10.1007/s11005-011-0515-8}{\textsf{pp.~375}},
\texttt{\arxivref{1012.3998}{1012.3998}}.

\chapitem{chap:N6}\label{chapN6}
T.~Klose,
\textit{``\hypref{chap:N6}{$\mathcal{N}$ = 6 Chern-Simons and Strings on $AdS_4 \times CP^3$}''},
\doiref{10.1007/s11005-011-0520-y}{\textsf{pp.~401}},
\texttt{\arxivref{1012.3999}{1012.3999}}.

\chapitem{chap:QCD}\label{chapQCD}
G.~Korchemsky,
\textit{``\hypref{chap:QCD}{Integrability in QCD and $\mathcal{N}$ $<$ 4 SYM}''},
\doiref{10.1007/s11005-011-0516-7}{\textsf{pp.~425}},
\texttt{\arxivref{1012.4000}{1012.4000}}.

\partitem{part:amplitudes}
\textbf{\hypref{part:amplitudes}{Scattering Amplitudes}}

\chapitem{chap:Amp}\label{chapAmp}
R.~Roiban,
\textit{``\hypref{chap:Amp}{Scattering Amplitudes -- a Brief Introduction}''},
\doiref{10.1007/s11005-011-0517-6}{\textsf{pp.~455}},
\texttt{\arxivref{1012.4001}{1012.4001}}.

\chapitem{chap:Dual}\label{chapDual}
J.~M.~Drummond,
\textit{``\hypref{chap:Dual}{Dual Superconformal Symmetry}''},
\doiref{10.1007/s11005-011-0519-4}{\textsf{pp.~481}},
\texttt{\arxivref{1012.4002}{1012.4002}}.

\chapitem{chap:TDual}\label{chapTDual}
L.~F.~Alday,
\textit{``\hypref{chap:TDual}{Scattering Amplitudes at Strong Coupling}''},
\doiref{10.1007/s11005-011-0518-5}{\textsf{pp.~507}},
\texttt{\arxivref{1012.4003}{1012.4003}}.

\partitem{part:algebra}
\textbf{\hypref{part:algebra}{Algebraic Aspects of Integrability}}

\chapitem{chap:Superconf}\label{chapSuperconf}
N.~Beisert,
\textit{``\hypref{chap:Superconf}{Superconformal Algebra}''},
\doiref{10.1007/s11005-011-0479-8}{\textsf{pp.~529}},
\texttt{\arxivref{1012.4004}{1012.4004}}.

\chapitem{chap:Yang}\label{chapYang}
A.~Torrielli,
\textit{``\hypref{chap:Yang}{Yangian Algebra}''},
\doiref{10.1007/s11005-011-0491-z}{\textsf{pp.~547}},
\texttt{\arxivref{1012.4005}{1012.4005}}.

\end{list}

\newpage

\fi

\section{Overview}
\label{sec:overview}

\begin{figure}[b!]\centering%
\includegraphics[width=0.88\textwidth]{FigOverview.mps}%
\caption{\ifarxiv 
Suggested order of study.\else 
An overview of the chapters and (some of) their connections.\fi}
\label{fig:parts}
\end{figure}

The chapters are grouped into six parts
representing the major topics and activities of this subject,
see \figref{fig:parts}.
In the first two \namedref{Parts}{part:gauge} \namedref{and}{part:string} we start by outlining 
the perturbative gauge and string theory setup.
Here we focus on down-to-earth quantum field theory calculations 
which yield the solid foundation in spectral data of local operators.
In the following \namedref{Part}{part:spectrum} we review the construction
of the spectrum by integrable methods. 
More than merely reproducing the previously obtained data,
this goes far beyond what could possibly be computed by conventional methods:
It can apparently predict the exact spectrum.
The next \namedref{Part}{part:beyond} summarises applications
of these methods to similar problems, beyond the spectrum, 
beyond planarity, beyond $\superN=4$ SYM or strings on $AdS_5\times S^5$.
Among these avenues is the application of integrability 
to scattering amplitudes; 
as this topic has grown into a larger subject 
we shall devote \namedref{Part}{part:amplitudes} to it.
The final \namedref{Part}{part:algebra} reviews classical and
quantum algebraic aspects of the models and of integrability.


\setcounter{secnumdepth}{2}
\def\thesection{\Roman{section}}

\newpage
\section{\texorpdfstring{$\mathcal{N}$}{N} = 4 Super Yang--Mills Theory}
\label{part:gauge}
\figinsertp{FigPartGauge.mps}

This part deals with the maximally supersymmetric Yang--Mills 
($\superN=4$ SYM) theory in four spacetime dimensions.
This model is a straight-forward quantum field theory. 
It uses the same types of particles and interactions 
that come to play in the Standard Model of particle physics.
However, the particle spectrum and the interactions are 
delicately balanced granting the model a host of unusual and unexpected features.
The best-known of these is exact (super)conformal symmetry at the quantum level.
A far less apparent feature is what this review collection is all about: 
integrability.

In this part we focus on the perturbative field theory,
typically expressed through Feynman diagrams.
The calculations are honest and reliable 
but they become tough as soon as one goes to higher loop orders. 
Integrability will only be discussed as far as it directly
concerns the gauge theory setup, 
i.e.~in the sense of conserved operators acting on a spin chain.
The full power of integrability will show up only in \namedref{Part}{part:spectrum}.

\subsection{Spin Chains in \texorpdfstring{$\mathcal{N}$}{N} = 4 SYM}
\label{chap:Chain}
\figinsert{FigChapChain.mps}

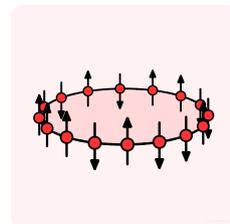
\chapref{Chain}
introduces the gauge theory, its local operators, 
and outlines how to compute the spectrum of their planar
one-loop anomalous dimensions.
It is explained how to map one-to-one
local operators to states of a certain quantum spin chain. 
The operator which measures the planar, 
one-loop anomalous dimensions corresponds
to the spin chain Hamiltonian in this picture. 
Importantly, this Hamiltonian is of the integrable kind,
and the planar model can be viewed as a generalisation of the Heisenberg spin chain.
This implies that its spectrum is solved
efficiently by the corresponding Bethe ansatz.
E.g.~a set of one-loop planar anomalous dimensions $\delta D$
is encoded in the solutions of the following set of Bethe equations
for the variables $u_k\in\Complex$ ($k=1,\ldots,M$)
\[
\lrbrk{\frac{u_k+\ihalf}{u_k-\ihalf}}^L=\mathop{\prod_{j=1}^M}_{j\neq k}\frac{u_k-u_j+i}{u_k-u_j-i}\,,
\qquad
1=\prod_{j=1}^M\frac{u_j+\ihalf}{u_j-\ihalf}\,,
\qquad
\delta D=\frac{\lambda}{8\pi^2}\sum_{j=1}^M\frac{1}{u_j^2+\quarter}\,.
\]
Finally the Chapter presents applications of the Bethe ansatz to sample problems.

\newpage

\subsection{The spectrum from perturbative gauge theory}
\label{chap:Higher}
\figinsert{FigChapHigher.mps}

The following \chapref{Higher}
reviews the computation of the spectrum of
anomalous dimensions at higher loops 
in perturbative gauge theory.
The calculation in terms of Feynman diagrams is firmly established, 
but just a handful orders takes you to the limit of what is generally possible.
Computer algebra and superspace techniques push the limit by a few orders.

In our case the results provide a valuable set of irrefutable data
which the integrable model approach must be able to reproduce
to show its viability. This comprises not only explicit energy eigenvalues, 
but also crucial data for the integrable system, such as the 
magnon dispersion relation and scattering matrix. 
Importantly, also the leading finite-size terms are accessible
in this approach, e.g.~the four-loop anomalous dimension
to the simplest non-trivial local operator reads
(cf.\ the above Bethe equation with $L=4,M=2$)
\[
\delta D=\frac{3\lambda}{4\pi^2}
-\frac{3\lambda^2}{16\pi^4}
+\frac{21\lambda^3}{256\pi^6}
-\frac{\bigbrk{78-18\zeta(3)+45\zeta(5)}\lambda^4}{2048\pi^8}
+\ldots
\]

\subsection{Long-range spin chains}
\label{chap:LR}
\figinsert{FigChapLR.mps}

The final \chapref{LR} of this part
reviews spin chain Hamiltonians originating 
in planar gauge theory at higher loops.
The one-loop Hamiltonian describes 
interactions between two neighbouring spins.
At higher loops the Hamiltonian is deformed
by interactions between several neighbouring spins, 
e.g.
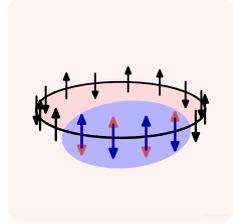
\[
H=\sum_{j=1}^L\lrbrk{\frac{\lambda}{8\pi^2}\bigbrk{1-P_{j,j+1}}+\frac{\lambda^2}{128\pi^4}\bigbrk{-3+4P_{j,j+1}-P_{j,j+2}}+\ldots}.
\]
Moreover, the Hamiltonian can dynamically add or remove spin sites! 
While integrable nearest-neighbour Hamiltonians 
have been studied in detail for a long time,
a better general understanding of long-range deformations 
was developed only recently.
Curiously, several well-known integrable spin chain models
make an appearance in this context, in particular, 
the Haldane--Shastry, Inozemtsev and Hubbard models.

\newpage
\section{IIB Superstrings on \texorpdfstring{$AdS_5\times S^5$}{AdS5xS5}}
\label{part:string}
\figinsertp{FigPartString.mps}

This part concerns IIB string theory on the maximally supersymmetric 
$AdS_5\times S^5$ background.
The string worldsheet model is a two-dimensional UV-finite quantum field theory. 
It is of the non-linear sigma model kind
with target space $AdS_5\times S^5$
and further possesses worldsheet diffeomorphisms. 
Also this model has a number of exceptional features,
such as kappa symmetry,
which make it a viable string theory on a stable background.
Somewhat less surprising than in gauge theory, 
this model is also integrable, 
a property shared by many two-dimensional sigma models on coset spaces. 

We outline how to extract spectral data from classical string solutions
with quantum corrections. 
There are many complications, 
such as non-linearity of the classical equations of motion,
lack of manifest supersymmetry and presence of constraints.
Again, integrability will help tremendously;
here we focus on string-specific aspects, 
and leave the more general applications to \namedref{Part}{part:spectrum}.

\subsection{Classical \texorpdfstring{$AdS_5\times S^5$}{AdS5xS5} string solutions}
\label{chap:Spinning}
\figinsert{FigChapSpinning.mps}

The first \chapref{Spinning} of this part
introduces the Green--Schwarz string 
on the curved spacetime $AdS_5\times S^5$.
For the classical spectrum only the
bosonic fields are relevant. 
To find exact solutions of the non-linear equations of motions,
one typically makes an ansatz for the shape of the string. 
Taking inspiration from spinning strings in flat space, 
one can for instance assume a geodesic rod spinning around some 
orthogonal axes. 
The equations of motion together with the Virasoro constraints 
dictate the local evolution, 
while boundary conditions quantise the string modes.
Next, the target space isometries give rise to conserved charges,
such as angular momenta and energy. 
These can be expressed in terms of the parameters
of the string solution.
E.g., a particular class of spinning strings on $AdS_3\times S^1\subset AdS_5\times S^5$ 
obeys the following relation
($\mathrm{K},\mathrm{E}$ are elliptic integrals)
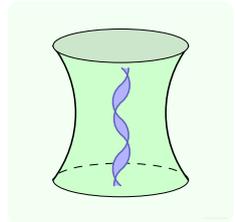
\[
\frac{S^2}{\bigbrk{\mathrm{K}(m)-\mathrm{E}(m)}^2}-\frac{J^2}{\mathrm{K}(m)^2}=16n^2T^2\,(1-m),
\qquad
\frac{J^2}{\mathrm{K}(m)^2}-\frac{E^2}{\mathrm{E}(m)^2}=16n^2T^2 \,m.
\]
Such relations can be used to express the energy $E$ as a function
of the angular momenta $J,S$, the string modes $n$ and the string tension $T$.%
\footnote{Note that complicated classes of solutions will require 
further internal parameters in addition to $m$.}

\newpage

\subsection{Quantum Strings in \texorpdfstring{$AdS_5\times S^5$}{AdS5xS5}}
\label{chap:QString}
\figinsert{FigChapQString.mps}

\chapref{QString}
continues with semiclassical quantisation of strings.
Here, one distinguishes between point-like and extended strings. 

Quantisation around point-like strings is the direct analogue 
of what is commonly done in flat space. 
The various modes of the string can be excited in quantised amounts,
and the string spectrum takes the form
\[
E-J =\sum_{k=1}^M
N_k\sqrt{1+\lambda n_k^2/J^2}
+\ldots,
\qquad
\sum_{k=1}^M N_k n_k=0.
\]
The main difference with flat space is that the modes interact,
adding non-trivial corrections to the spectrum.
These corrections can be computed in terms of a scattering problem on
the worldsheet.

Quantisation around extended string solutions is far less trivial:
The spectrum of fluctuations now crucially depends on the classical solution. 
Another effect is that the energy of the classical string receives quantum corrections
due to vacuum energies of the string modes.

\subsection{Sigma Model, Gauge Fixing}
\label{chap:Sigma}
\figinsert{FigChapSigma.mps}

Spheres and anti-de-Sitter spacetimes are symmetric cosets.
\chapref{Sigma} presents the formulation of
the string worldsheet as a two-dimensional coset space sigma model
of the target space isometry supergroup. 
In particular, integrability finds a simple formulation 
in a family of flat connections $A(z)$ on the worldsheet
and its holonomy $M(z)$ around the closed worldsheet
\[
dA(z)+A(z)\wedge A(z)=0,
\qquad
M(z)=\mathrm{P}\exp\oint A(z).
\]
Series expansion of $M(z)$ in the spectral parameter $z$
leads to an infinite tower of charges
extending the isometries 
to an infinite-dimensional algebra.

Proper treatment of symmetries and integrability
towards a canonical quantisation 
requires a Hamiltonian formulation.
Here the major complications are the presence of first and second class constraints 
due to worldsheet diffeomorphisms and kappa symmetry.
Finally, one encounters notorious ambiguities in deriving
the algebra of conserved charges.

\newpage

\subsection{The Spectral Curve}
\label{chap:Curve}
\figinsert{FigChapCurve.mps}

In the final \chapref{Curve} on strings,
the flat connection is applied to the construction 
of the (semi)classical string spectrum.
The eigenvalues $e^{ip_k(z)}$ of the monodromy $M(z)$ are integrals of motion. 
As functions of complex $z$ they define a spectral curve 
for each classical solution.
Instead of studying explicit classical solutions 
we can now study abstract spectral curves.
Besides containing all the spectral information,
they offer a neat physical picture:
String modes correspond to handles of the Riemann surface,
and each handle has two associated moduli: the mode number
$n_k$ and an amplitude $\alpha_k$. They can be extracted easily 
as periods of the curve
\[
\oint_{A_k} dp=0,\qquad
\frac{1}{2\pi}\oint_{B_k} dp=n_k,\qquad
\frac{\sqrt{\lambda}}{4\pi^2i}\oint_{A_k}\frac{1+z^4}{1-z^4}\,dp=N_k.
\]
Note that quantisation replaces the amplitude by an integer excitation number $N_k$
thus providing access to the semiclassical spectrum of fluctuation modes.

\section{Solving the AdS/CFT Spectrum}
\label{part:spectrum}
\figinsertp{FigPartSpectrum.mps}

Armed with some basic knowledge of the 
relevant structures in gauge and string theory
(as well as an unconditional belief in the applicability 
of integrable structures to this problem)
we aim to solve the planar spectrum
in this part.

The starting point is that in both models 
there is a one-dimensional space (spin chain, string) 
on which some particles (magnons, excitations) can propagate. 
By virtue of symmetry and integrability one can derive how they scatter,
at all couplings and in all directions. 
Taking periodicity into account properly, one arrives at a complete
and exact description of the spectrum.
For certain states this program was carried out, 
and all results are in complete agreement with explicit
calculations in perturbative gauge or string theory
(as far as they are available). 
Yet, the results from the integrable system approach go far beyond 
what is otherwise possible in QFT: They provide a window to finite coupling $\lambda$!

There are several proposals of how to formulate these equations ---
through an algebraic system or through integral equations.
However, it is commonly believed that a reasonably simple and generally usable 
form for such equations has not yet been found (let alone proved).

\newpage

\subsection{Bethe Ans\"atze and the R-Matrix Formalism}
\label{chap:ABA}
\figinsert{FigChapABA.mps}

As a warm-up exercise and to gather experience, 
\chapref{ABA} 
solves one of the oldest quantum mechanical systems --- 
the Heisenberg spin chain.
This is done along the lines of Bethe's original work, 
using a factorised magnon scattering picture, 
but also in several variants of the Bethe ansatz. 
This introduces us to ubiquitous concepts
of integrable systems
such as R-matrices, transfer matrices
and the famous Yang--Baxter equation
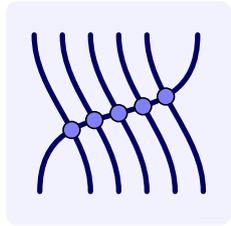
\[
R_{12} R_{13} R_{23} = R_{23} R_{13} R_{12}.
\]
	The chapter ends by sketching a promising novel method for constructing the
so-called Baxter Q-operators, allowing to surpass the Bethe ansatz technique.

\subsection{Exact world-sheet S-matrix}
\label{chap:SMat}
\figinsert{FigChapSMat.mps}

Even though little is known about gauge or string theory at finite coupling,
the magnon scattering pictures and symmetries 
qualitatively agree for weak and strong coupling. 
Under the assumption that they remain valid at intermediate couplings,
\chapref{SMat} describes how to make use of symmetry 
to determine all the relevant quantities: 
Both, the magnon dispersion relation
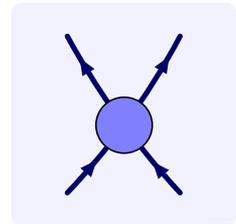
\[
e(p)=\sqrt{1+\frac{\lambda}{\pi^2}\sin^2(\half p)}
\]
and the 16-flavour scattering matrix 
are almost completely determined through representation theory
of an extended $\alg{psu}(2|2)$ superalgebra.
Integrability then ensures factorised scattering,
and determines the spectrum on sufficiently long chains or strings
through the asymptotic Bethe equations.


\subsection{The dressing factor}
\label{chap:SProp}
\figinsert{FigChapSProp.mps}

Symmetry alone cannot predict an overall phase factor 
of the scattering matrix
which is nevertheless crucial for the spectrum. 
Several other desirable properties of factorised scattering systems,
such as unitarity, crossing and fusion,
constrain its form
\[
S^0_{12}S^0_{1\bar 2} = f_{12}.
\]
\chapref{SProp} presents this crossing equation
and its solution -- the so-called dressing phase. 
It has a host of interesting analytic properties
relating to the physics of the model under discussion.

\newpage

\nextparspace
\subsection{Twist states and the cusp anomalous dimension}
\label{chap:Twist}
\figinsert[\figinserttwoline]{FigChapTwist.mps}

The asymptotic Bethe equations predict the spectrum 
up to finite-size corrections. 
In \chapref{Twist} we apply them 
to the interesting class of twist states.
These are ideally suited for testing purposes 
because a lot of solid spectral data are known 
from perturbative gauge and string theory.
They also have an interesting dependence on their spin $j$,
in terms of generalised harmonic sums of fixed degree.

Importantly, in the large spin limit, finite-size corrections 
turn out to be suppressed.
The Bethe equations reduce to an integral equation 
to predict the exact cusp dimension (and generalisations).
The latter turns out to interpolate smoothly between weak and strong coupling
in full agreement with perturbative data
\[
D\indup{cusp}=
\frac{\lambda}{2\pi^2}
-\frac{\lambda^2}{96\pi^2}
+\frac{11\lambda^3}{23040\pi^2}
+\ldots
=
\frac{\sqrt{\lambda}}{\pi}-\frac{3\log 2}{\pi}-\frac{\beta(2)}{\pi\sqrt{\lambda}}+\ldots
\,.
\]
%

\subsection{L\"uscher corrections}
\label{chap:Luescher}
\figinsert{FigChapLuescher.mps}

For generic states, however, finite-size corrections 
are required to get agreement with gauge and string theory.
\chapref{Luescher} explains how to apply 
L\"uscher terms to determine these:
On a closed worldsheet there are virtual particles propagating 
in the spatial direction in non-trivial loops around the string.
When they interact with physical excitations, 
they give rise to non-trivial energy shifts
($q_j$ and $p_k$ are virtual and real particle momenta, respectively)
\[
\delta E=-\sum\nolimits_j\int \frac{dq_j}{2\pi} \,
e^{-L \tilde{e_j}(q_j)} \STr\nolimits_j\prod\nolimits_k S_{jk}(q_j,p_k)
.
\]
%

\subsection{Thermodynamic Bethe Ansatz}
\label{chap:TBA}
\figinsert{FigChapTBA.mps}

Although finite-size corrections appear under control,
it is clearly desirable to find equations to 
determine the exact spectrum in one go. 
\chapref{TBA} describes the thermodynamic Bethe
ansatz approach based on the following idea:
Consider the string worldsheet at finite temperature
with Wick rotated time. 
It has the topology of a torus
of radius $R$ and time period $L$. 
We are primarily interested in the zero temperature limit
where time is decompactified. 
Now the torus partition function can be evaluated 
in the mirror theory where the periods are exchanged 
\[
Z(R,L)=\tilde Z(L,R).
\]
Then, instead of time, we can decompactify the radius.
Conveniently, the asymptotic Bethe equations become exact,
and eventually predict the finite-size spectrum.

\newpage
\subsection{Hirota Dynamics for Quantum Integrability}
\label{chap:Trans}
\figinsert{FigChapTrans.mps}

\chapref{Trans} presents a equivalent proposal 
for the finite-size spectrum
based on the conserved charges of an integrable model.
The latter are typically packaged into transfer matrix eigenvalues $T(u)$.
These exist in various instances
which obey intricate relations, 
such as the discrete Hirota equation
(also known as the Y-system for equivalent quantities $Y_{a,s}(u)$)
\[
T_{a,s}(u+\ihalf)T_{a,s}(u-\ihalf)=
T_{a+1,s}(u)T_{a-1,s}(u)
+T_{a,s+1}(u)T_{a,s-1}(u).
\]
Similarly to \namedref{Chapter}{chap:Curve},
one can start from these equations, 
subject to suitable boundary conditions, 
and predict the spectrum at finite coupling.

\section{Further Applications of Integrability}
\label{part:beyond}
\figinsertp{FigPartBeyond.mps}

For the sake of a clear presentation the previous parts focused 
on one particular application of integrability in AdS/CFT: 
Solving the exact planar spectrum of $\superN=4$ supersymmetric 
Yang--Mills theory or equivalently of IIB string theory on $AdS_5\times S^5$. 
While this topic has been at the centre of attention, 
many investigations have dealt with extending the applications of integrability
to other observables beyond the planar spectrum and to more general models. 
This part and the following try to give an overview of these developments.

\subsection{Aspects of Non-planarity}
\label{chap:Observ}
\figinsert{FigChapObserv.mps}

Integrability predicts the planar spectrum accurately and with minimum effort. 
It would be desirable to extend the applications of integrability to non-planar corrections
because, e.g., in QCD $N\indup{c}=3$ rather than $N\indup{c}=\infty$.
For the spectrum, these are interactions 
where the spin chain or the string splits up and recombines
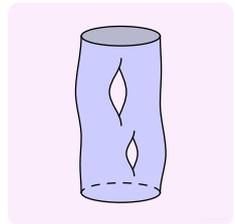
\[
H=H_0+\frac{1}{N\indup{c}}\,(H_+ +H_-)+\ldots\,.
\]
They result in a string worldsheet of higher genus or with more than two punctures.

\chapref{Observ} reviews the available results 
on higher-genus corrections, higher-point functions as well as 
supersymmetric Wilson loops in the AdS/CFT context.
It is shown that most of the basic constructions of integrability 
do not work in the non-planar setup.

\newpage
\ifarxiv\nextparspace\fi
\subsection{Deformations, Orbifolds and Open Boundaries}
\label{chap:Deform}
\figinsert[\ifarxiv\figinserttwoline\else\figinsertoneline\fi]{FigChapDeform.mps}

There exist many deformations of $\superN=4$ SYM which preserve
some or the other property, e.g.\ by deforming the ($\superN=1$) superpotential
\[
\int d^4x\, d^4\theta\Tr\,\bigbrk{e^{i\beta} XYZ-e^{-i\beta} ZYX}.
\]
It is natural to find out under which conditions 
integrability can survive. 
\chapref{Deform}
reviews such superconformal deformations of $\superN=4$ SYM and shows how the
methods of integrability can be adjusted to these cases.
It turns out that these merely deform the boundary conditions of 
the integrable model by introducing additional phases into the Bethe equations
(in the spin chain context this has a similar effect as turning on a magnetic field). 
Different boundary conditions can also arise from looking at
other corners of the spectrum or at different observables;
this is another topic of the present chapter.

\nextparspace
\subsection{\texorpdfstring{$\mathcal{N}$}{N} = 6 Chern-Simons and Strings on \texorpdfstring{\\}{}\texorpdfstring{$AdS_4 \times CP^3$}{AdS4xCP3}}
\label{chap:N6}
\figinsert[\figinserttwoline]{FigChapN6.mps}

Recently a quantum field theory in three dimensions 
was discovered which behaves in many respects like $\mathcal{N}=4$ SYM ---
$\mathcal{N}=6$ supersymmetric Chern--Simons-matter theory
\[
S=\frac{k}{4\pi}\int \Tr\,\bigbrk{A\wedge dA+\sfrac{2}{3}A\wedge A\wedge A+\ldots}.
\]
It is exactly superconformal at the quantum level,
and there is an AdS/CFT dual string theory ---
IIA superstrings on the $AdS_4\times CP^3$ background.
Importantly, there exists a large-$N\indup{c}$ limit,
in which the model becomes integrable. 
\chapref{N6} reviews integrability 
in this AdS$_4$/CFT$_3$ correspondence. 
While being largely analogous to the constructions in the previous parts, 
there are several noteworthy differences in the application of integrable methods:
For instance, here the spin representation alternates between the sites.

\subsection{Integrability in QCD and \texorpdfstring{$\mathcal{N}$}{N} \texorpdfstring{$<$}{<} 4 SYM}
\label{chap:QCD}
\figinsert{FigChapQCD.mps}

Similar integrable structures were known to exist 
in more general gauge theories
long before the exploitation of integrability 
in $\superN=4$ SYM. 
\chapref{QCD} introduces 
evolution equations for high-energy scattering (BFKL) and
scaling of quasi-partonic operators
in connection to deep inelastic scattering (DGLAP). 
To some extent these take the form of integrable Hamiltonians 
with $\alg{sl}(2|\superN)$ symmetry
($J_{12}$ is the two-particle spin operator and
$\mathrm{\Psi}$ is the digamma function)
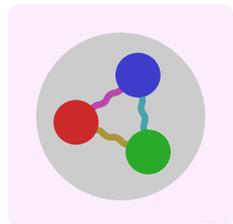
\[
H_{12}\simeq 
\mathrm{\Psi}(J_{12})-\mathrm{\Psi}(1).
\]
Its eigenvalues determine the scaling of certain 
hadronic structure functions 
and control the energy dependence of 
scattering amplitudes in the high-energy (Regge) limit.

\newpage
\nextparspace[3.0cm]
\section{\texorpdfstring{\!\!}{}Integrability\texorpdfstring{\!}{} for\texorpdfstring{\!}{} Scattering\texorpdfstring{\!}{} Amplitudes}
\label{part:amplitudes}
\figinsertp{FigPartAmplitudes.mps}

The most conservative application of quantum field theories
is to compute scattering cross sections 
(to be compared to particle scattering experiments). 
With old blades sharpened and new ones developed,
the charted territory of tree and loop scattering amplitudes 
in $\superN=4$ SYM has increased dramatically,
see e.g.\ the recent reviews \cite{Dixon:2008tu,Alday:2008yw,Henn:2009bd,Wolf:2010av,Drummond:2010ep}
and the special issue \cite{Roiban:2011zz}.
It was soon noticed that something special was going on in the planar limit
which makes amplitudes much simpler than originally thought.
It does not take much imagination to conjecture a connection to integrability.
This part reviews scattering amplitudes
and what integrability implies in this context. 
This topic is under active investigation,
many advances have been and are being made, 
but a lot remains to be understood.
Here, one can expect that integrability 
will enable a similarly simple solution 
as in the case of the planar spectrum. 

\subsection{Scattering Amplitudes -- a Brief Introduction}
\label{chap:Amp}
\figinsert{FigChapAmp.mps}

\chapref{Amp} gives an introduction into the topic
of scattering amplitudes in $\superN=4$ SYM. 
First of all, the spinor-helicity formalism and colour-ordering scheme
strips the combinatorial structure and leaves plain functions. 
For instance, an essential part of an $n$-particle amplitude simply reads
($\langle kj\rangle$ is a Lorentz-invariant constructed from the momenta of particles $k$ and $j$)
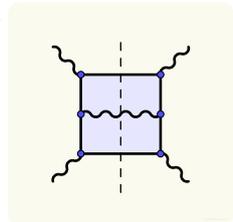
\[
A\supup{MHV}_n=\frac{\delta^4(P)\,\delta^8(Q)}{
\langle 12\rangle \langle 23\rangle \cdots \langle n1\rangle}\,.
\]
The S-matrix displays a host of useful analyticity properties
related to unitarity. 
These can be used to reconstruct tree and loop amplitudes from scratch,
which is typically far more efficient than using Feynman diagrams
following from the Lagrangian description.

\subsection{Dual Superconformal Symmetry}
\label{chap:Dual}
\figinsert{FigChapDual.mps}

\chapref{Dual} reviews simplifications
found in planar scattering. 
It turns out that the underlying scalar integrals are of a special form
which hints at conformal symmetry in a dual space. 
Indeed, the amplitudes obey a dual superconformal symmetry 
in addition to the conventional one. 
The two sets of conformal symmetries close onto
an infinite-dimensional algebra 
which is at the heart of integrability --- the Yangian.

This symmetry helps to determine all (tree) amplitudes,
by means of recursion or through a Grassmannian integral
($C$ is a $k\times n$ matrix, 
$M_j$ are its $k\times k$ minors of consecutive columns,
and $Z$ are $4|4$ twistors encoding the momenta of the $n$ legs)
\[
A\supup{tree}_{n,k}(Z)=\int \frac{d^{k(n-k)}C\, \delta^{k(4|4)}(CZ)}{M_1\cdots M_n}\,.
\]
%

\subsection{Scattering Amplitudes at Strong Coupling}
\label{chap:TDual}
\figinsert{FigChapTDual.mps}

\chapref{TDual} discusses the string dual 
of scattering amplitudes.
Here it makes sense to transform particle momenta 
to distances by means of a T-duality. 
At strong coupling an amplitude is then dominated 
by the minimal area of a string worldsheet 
ending on a light-like polygonal contour 
on the $AdS_5$ boundary
(the previous \namedref{Chapter}{chap:Dual}
provides evidence in favour of a general relation 
between amplitudes and light-like polygonal Wilson loops).
Such minimal areas can be computed efficiently by 
integrable means bypassing the determination 
of the complicated shape of the worldsheet
(cf.\ \namedref{Chapter}{chap:Luescher})
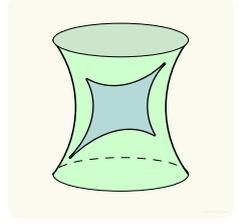
\[
A\indup{reg}=\sum_k\int\frac{d\theta\, m_k \cosh \theta}{2\pi} 
\log\bigbrk{1+Y_k(\theta)}.
\]

\section{Algebraic Aspects of Integrability}
\label{part:algebra}
\figinsertp{FigPartAlgebra.mps}

Integrability can be viewed as a symmetry.
In most cases it enhances an obvious, finite-dimensional symmetry of a physical system 
to a hidden, infinite-dimensional algebra. 
The extended symmetry then imposes a large number of constraints 
onto the system
which determine the dynamics (almost) completely,
but without making it trivial.
Many of the properties and methods that come to use in integrable systems 
find a mathematical formulation in terms of quantum algebra. 
Often this does not help immediately in computing particular physical observables, 
one of the main objectives of the previous parts.
Rather, it can give a deeper understanding of how the model's
integrability works, with a view to finding rigorous proofs
for the applicability of the (well-tested) proposals.

This final part of the review presents the symmetries relevant 
to our gauge and string theory problem. 
These are the Lie supergroup $\grp{PSU}(2,2|4)$ as the obvious symmetry
and its Yangian algebra to encode integrability.


\subsection{Superconformal Algebra}
\label{chap:Superconf}
\figinsert{FigChapSuperconf.mps}

The Lie superalgebra $\alg{psu}(2,2|4)$
is generated by $8\times 8$ supermatrices (in $2|4|2$ grading)
\[
J=\matr{c|c|c}{L&-iQ&P\vphantom{\hat{\bar{Q}}}\\\hline S&R&\bar Q\vphantom{\hat{\bar{Q}}}\\\hline K&-i\bar S&\bar L\vphantom{\hat{\bar{Q}}}},
\]
subject to suitable constraints, projections and hermiticity conditions.
It serves as the spacetime superconformal symmetry in gauge theory 
as well as the target space isometries of the dual string theory.

\chapref{Superconf} summarises some well-known facts and results
for this algebra. It also explains how the algebra applies to the 
gauge and string theory setup. 
The chapter is not so much related to integrability itself, 
it can rather be understood as an appendix to many of
the other chapters when it comes to the basics of symmetry.

\newpage
\subsection{Yangian Algebra}
\label{chap:Yang}
\figinsert{FigChapYang.mps}

In physics one is used to the concept of locally and homogeneously acting symmetries.
\chapref{Yang} introduces the Yangian algebra 
whose non-local action is encoded by the coproduct
\[
\mathrm{\Delta}(Y^A)=
Y^A\otimes 1
+1\otimes Y^A
+f^A_{BC} J^B\otimes J^C.
\]
For instance, such a non-local action permits a scattering matrix
which is fully determined by the algebra while still being non-trivial. 
The scattering matrix becomes 
a natural intertwining object of the Yangian, its R-matrix.
It enjoys a host of useful properties 
which eventually make the physical system tractable.

\setcounter{secnumdepth}{0}

\section{Acknowledgements}

N.~Beisert thanks T.\ Bargheer, M.\ de Leeuw and L.\ Lipatov for helpful comments on the manuscript.

The work of C.\ Ahn was supported in part by KRF-2007-313-C00150
and WCU grant R32-2008-000-10130-0 (CA).
R.\ Janik was supported in part by Polish science funds as a research project
N~N202~105136 (2009--2011).
The work of C.\ Kristjansen and K.\ Zoubos was supported by FNU through
grant number 272-08-0329.
The work of R.\ I.\ Nepomechie was supported in part 
by the National Science Foundation under Grants PHY-0554821 and PHY-0854366 (RN).
The work of A.\ Torrielli has been partially supported 
by EPSRC through the grant no.\ EP/H000054/1.

\phantomsection
\addcontentsline{toc}{section}{\refname}
\bibliographystyle{nb}
\bibliography{intads,chapters}

\end{document}